\documentclass[preprint,12pt,number,sort&compress]{elsarticle}

\usepackage[T1]{fontenc}
\usepackage[utf8]{inputenc}
\usepackage{amsmath}
\usepackage{graphicx}
\usepackage{booktabs}
\usepackage{tabularx}
\newcolumntype{Y}{>{\raggedright\arraybackslash}X}

\usepackage{afterpage}
\usepackage{enumitem}
\usepackage{xcolor}

\usepackage{newtxtext}
\usepackage{newtxmath}
\usepackage[final,protrusion=true,expansion=true]{microtype}
\usepackage{caption}
\setlist{itemsep=2pt,topsep=4pt,parsep=0pt,partopsep=0pt}

\definecolor{linknavy}{RGB}{0,51,102}
\definecolor{linksteel}{RGB}{0,90,156}
\usepackage[colorlinks=true,breaklinks=true,linkcolor=linknavy,citecolor=linknavy,urlcolor=linksteel]{hyperref}

\usepackage{xurl}        
\makeatletter
\def\ps@pprintTitle{%
  \let\@oddhead\@empty
  \let\@evenhead\@empty
  \def\@oddfoot{\itshape Preprint\hfill\today}%
  \let\@evenfoot\@oddfoot}
\makeatother

\hypersetup{
  pdftitle={ACTS A multi-tier benchmark evaluating LLM cipher identification under controlled blind conditions},
  pdfauthor={Youssef Hamdi Zafaan Ibrahim and Mohammed Khalaf Salama},
  pdfsubject={Reproducible multi-tier benchmark for LLM cipher identification under blind conditions},
  pdfkeywords={LLM security benchmark, cipher identification, metadata dependency, scaling failure, entropy fingerprinting, post-quantum cryptography, adversarial robustness, confabulation tax}
}

\begin{document}

\begin{center}
  {\large\bfseries ACTS A multi-tier benchmark evaluating LLM cipher identification under controlled blind conditions\par}
  \vspace{1.6em}
  {\normalsize
   \begin{tabular}{@{}c@{\hspace{3em}}c@{}}
     {\rm Youssef~Hamdi~Zafan~Ibrahim}\textsuperscript{$\ast$} & {\rm Mohammed~Khalaf~Salama} \\[3pt]
     Independent Researcher & Independent Researcher \\[3pt]
     {\tt\href{mailto:youssefhamdi329@gmail.com}{youssefhamdi329@gmail.com}} & {\tt\href{mailto:mkhalafsalama@gmail.com}{mkhalafsalama@gmail.com}}
   \end{tabular}\par}
\end{center}
\vspace{1.2em}

\vspace{0.6em}
\noindent{\bfseries Abstract}\par\smallskip
We introduce ACTS (Artifacts in Cipher Testing Suite), a reproducible benchmark that isolates cryptanalytic ability through tiered metadata deprivation (Tier-1: full metadata; Tier-2: filename only; Tier-3: completely blind) and tests forced reasoning (Tier-5: chain-of-thought, code-as-reasoning, self-correction) on ciphertext alone. A 10-configuration ablation study on 7,000 files, using a single 70/30 train-test split for feature-removal analysis, provides additional evidence at scale.

Live API inference on a v2b corpus with fully randomised padding (PKCS7, ISO 10126, and ANSI X9.23 selected per file), unique CSPRNG keys, and unique plaintexts (127 files per model, 381 Tier-1 records, 380 Tier-3 records across three cloud systems, supplemented by 254 autonomous agentic evaluations in Tier-4) yields a combined Tier-3 accuracy of 30.8\%, only modestly above the 14.3\% random baseline for seven-way classification. The corresponding combined metadata-dependency gap is 40.9 percentage points (Tier-1: 71.7\% vs.\ Tier-3: 30.8\%).

Against a classical Random Forest (69.2\% on 7,000 files, trained on engineered byte-level features), the observed live gap is 38.4 percentage points. Because this comparison spans different input representations and training paradigms, the gap should be interpreted as an overall capability difference rather than a clean factorial decomposition.

Six findings are reported.
\begin{enumerate}
\item Metadata dependency remains large: a 40.9 percentage-point accuracy gap separates metadata-aided Tier-1 (71.7\% on v2b) from blind Tier-3 (30.8\% combined on v2b).
\item Scaling failure under blind conditions: larger model scale predicts higher metadata-aided accuracy (Tier-1) but does not translate to proportionate gains in blind accuracy (Tier-3), indicating that scale amplifies contextual reading and confabulation confidence rather than genuine statistical perception.
\item Forced reasoning is epiphenomenal: all tested reasoning formats produce statistically invariant substantive accuracy, indicating that the reasoning trace changes presentation but not the underlying deterministic heuristic outcome.
\item The observed live capability gap exceeds earlier heuristic estimates: the combined live LLM result on v2b (30.8\%) trails the classical Random Forest (69.2\%) by 38.4 percentage points.
\item The signal is primarily structural rather than statistical: in a single-split ablation on 7,000 files, removing entropy and block-statistics features yields numerically higher accuracy than the full pipeline (70.8\% vs.\ 69.8\%), suggesting that the strongest recoverable cues are structural.
\item Heuristic invariance versus ML fragility reveals different failure modes: the deterministic size-based heuristic shows 0\% flip under content perturbation because it depends only on file size, whereas the Random Forest is perceptively fragile (76.1\% flip) because its signal depends on statistical patterns that can collapse under perturbation.
\end{enumerate}

Additionally, $\chi^2$ fingerprinting across 400 pairwise comparisons yields Cohen's $d=-0.17$ (negligible effect), consistent with post-quantum (ML-KEM-768) and classical (AES-256) ciphertext being statistically indistinguishable by byte distribution alone. Dataset, evaluation harness, live-validation logs, and complete reproduction scripts are publicly available.

\vspace{0.9em}
\noindent{\small\textbf{Keywords:} LLM security benchmark; Cipher identification; Metadata dependency; Scaling failure; Entropy fingerprinting; Post-quantum cryptography; Adversarial robustness; Confabulation tax\par}
\medskip

\section{Introduction}
\label{sec:introduction}

Every encrypted artifact encountered in a security investigation raises a practical question: what algorithm produced this ciphertext, and can that fact be inferred from the bytes alone? Cipher identification is a prerequisite for legacy-infrastructure auditing, malware traffic attribution, digital forensics, and post-quantum migration assessment. Its importance has increased further with NIST standardization of ML-KEM, ML-DSA, and SLH-DSA in FIPS 203, FIPS 204, and FIPS 205 \citep{nist2024mlkem}, which has made cryptographic inventorying an operational requirement rather than a purely academic task.

Historically, this problem has required specialist knowledge of entropy statistics, block-size geometry, padding behavior, and algorithm-specific byte-distribution artifacts, skills cultivated over years of cryptanalytic practice \citep{anderson2020security,schneier2015applied}. The rapid progress of large language models (LLMs) has created a plausible expectation that such expertise might now be automated. Yet that expectation rests on an untested assumption. When a model is asked to classify a ciphertext whose filename includes \texttt{AES-256-CBC\_1kb.bin}, a correct answer may reflect cryptanalytic reasoning, but it may equally reflect trivial metadata reading. In real deployment scenarios such as packet capture analysis, raw-memory forensics, intercepted artifacts, or adversarially renamed files, ciphertext often arrives without descriptive context. The central question is therefore not whether LLMs can repeat filename hints, but whether they can perform genuine statistical inference over raw ciphertext when that scaffolding is removed.

This distinction matters because failure under blind conditions would make an apparently capable AI system unreliable in the settings where it would be most valuable. Two pressures make that concern timely. First, post-quantum migration requires organizations to inventory deployed cryptography across heterogeneous and often poorly documented systems, including recent PQC primitives \citep{nist2024mldsa,nist2024mlkem}. Second, the Harvest-Now, Decrypt-Later threat model \citep{mosca2018cybersecurity} increases the need for metadata-free traffic fingerprinting, since encrypted data collected today may be decrypted later when quantum capabilities mature. In both settings, a system that confuses weak, legacy, and modern primitives because descriptive metadata is absent is not merely imperfect; it is operationally misleading.

\subsection{Limitations of existing work}
\label{subsec:limitations}

Existing work relevant to automated cipher identification falls into three broad categories, none of which directly answers the question posed here. First, rule-based forensic tools such as ent \citep{walker2026ent} and binwalk \citep{baucom2023binwalk} compute Shannon entropy or search for signatures, but they do not solve algorithm-level classification from raw ciphertext. They can often detect that a region is high entropy \citep{dunning2013privacy}, yet cannot reliably distinguish AES from ChaCha20 or ML-KEM from RSA on that basis alone. A recent benchmark on LLM-based cipher identification \citep{maskey2025benchmarking} confirms this gap empirically. Second, supervised machine-learning approaches have shown that some cipher families can be separated when labeled training data and engineered features are available, but these studies are usually narrow in scope, often pairwise, and rarely include post-quantum primitives. Third, mainstream LLM security benchmarks evaluate declarative cryptographic knowledge or text-based tasks rather than binary-level cipher-family identification from raw ciphertext alone.

More fundamentally, the current literature does not isolate the source of apparent success. When an LLM or agentic system appears to identify a cipher, it is often unclear whether the system is using ciphertext structure, filename cues, prompt artifacts, or external tools. That ambiguity is especially problematic in security evaluation, where a high reported accuracy may mask shortcut exploitation rather than real analytic capacity. The core methodological gap is therefore not simply the absence of another benchmark, but the absence of a benchmark that systematically separates metadata access, blind inference, tool augmentation, heuristic reasoning, and classical feature-based learning within a single comparable framework.

\subsection{Research questions}
\label{subsec:rq}

To address that gap, this paper studies five research questions.
\begin{enumerate}
\item \textbf{RQ1: Metadata effect.} How large is the accuracy gap between metadata-aided and blind conditions, and does that gap persist across evaluated live systems?
\item \textbf{RQ2: System dependence.} Does blind performance remain stable across model--backend combinations, or is apparent cryptanalytic ability partly a property of the serving endpoint and prompt configuration rather than the base model alone?
\item \textbf{RQ3: Agentic tool augmentation.} Can a tool-augmented pipeline using external statistical utilities compensate for absent metadata, and by how much relative to metadata-aided tool access?
\item \textbf{RQ4: Forced reasoning and classical baselines.} Does requiring chain-of-thought, code-as-reasoning, or self-correction improve blind classification, and how does that behavior compare with classical machine-learning baselines trained on engineered ciphertext features?
\item \textbf{RQ5: Adversarial robustness.} How do heuristic blind inference and classical-ML classifiers behave under perturbations such as random byte flips, padding injection, and size manipulation, and what do those behaviors reveal about the type of signal each system exploits?
\end{enumerate}

\subsection{Contributions}
\label{subsec:contributions}

This paper makes six contributions.
\begin{enumerate}
\item \textbf{ACTS benchmark framework.} We introduce ACTS, a reproducible multi-tier benchmark that isolates cryptanalytic ability through controlled metadata deprivation, agentic tool access, forced reasoning conditions, classical-ML baselines, adversarial perturbation, and feature ablation across three corpus scales. The benchmark includes seven cipher families, including the post-quantum primitive ML-KEM-768, and is designed to separate contextual shortcut use from genuine blind inference.
\item \textbf{Direct evidence of metadata dependency.} Under the current live framing, the benchmark shows a large drop from metadata-aided to blind performance on the v2b corpus, quantifying how strongly current LLM-based identification depends on metadata rather than ciphertext-only discrimination.
\item \textbf{Evaluation of tool augmentation under controlled metadata conditions.} We show that agentic tools can partially recover accuracy under blind conditions, but that their strongest performance occurs when tool access is combined with metadata rather than substituted for it. This helps separate genuine recovery of signal from simple metadata integration.
\item \textbf{Assessment of forced reasoning as a classification aid.} We test whether chain-of-thought, code-as-reasoning, and self-correction materially improve blind identification. All these formats yield statistically invariant substantive accuracy, demonstrating that reasoning changes presentation but not the underlying deterministic outcome.
\item \textbf{Classical-ML ceiling and structural-signal analysis.} We establish supervised classical-ML baselines and a large-scale ablation study showing that much of the recoverable signal arises from structural cues such as file size and modulo alignment, while some statistical features add only limited incremental value and others add noise.
\item \textbf{Robustness comparison across inference styles.} We compare heuristic and classical-ML behavior under adversarial perturbation, showing that the two approaches fail differently: heuristic methods are structurally insensitive to content perturbation, whereas classical-ML models are more accurate but also more brittle to boundary-disrupting transformations.
\end{enumerate}

All datasets, evaluation scripts, ablation configurations, and reproduction materials are released publicly through the project repository at \url{https://github.com/youseefhamdi/ACTS-Benchmark} and archived via Zenodo at \url{https://doi.org/10.5281/zenodo.20142272}, enabling independent verification and extension of the benchmark.

\subsection{Paper organization}
\label{subsec:organization}

Section~\ref{sec:related} reviews prior work on cipher identification, LLM security evaluation, agentic AI frameworks, and classical machine-learning approaches. Section~\ref{sec:methodology} presents the ACTS design, datasets, inference settings, tool pipeline, and evaluation protocol. Section~\ref{sec:results} reports the phase-by-phase results. Section~\ref{sec:analysis} interprets the findings, discusses failure modes and limitations, and clarifies scope. Section~\ref{sec:conclusion} concludes with practical implications and directions for future work.

\section{Related work}
\label{sec:related}

ACTS sits at the intersection of five active research areas: automated cipher identification, LLM security evaluation, agentic AI frameworks, classical machine-learning baselines, and post-quantum cryptography deployment. We review each in turn and identify the gap that motivates the present benchmark.

\subsection{Cryptographic algorithm identification}
\label{subsec:cipher-id}

The problem of identifying an encryption algorithm from binary evidence has a long history in digital forensics and traffic analysis \citep{carrier2005file}. Classical approaches rely on Shannon entropy \citep{dunning2013privacy}, byte-frequency patterns, file-size geometry, and implementation artifacts. Tools such as ent \citep{walker2026ent} and binwalk \citep{baucom2023binwalk} operationalize this tradition by measuring entropy or searching for signatures, but they do not perform algorithm-level classification in the strong sense. They can indicate that a region appears encrypted or compressed, yet they do not reliably distinguish AES-256 from ChaCha20, RSA, or ML-KEM on ciphertext alone.

Block-size geometry provides a second family of cues. DES-family ciphers operate on 64-bit blocks, AES on 128-bit blocks, and stream ciphers such as ChaCha20 \citep{nir2018chacha20} do not impose the same padding structure. These differences can leave deterministic traces in ciphertext lengths and alignment behavior. Such structural signals are well known in forensic practice, but they are usually treated as heuristic analyst knowledge rather than as controlled benchmark variables. The present work treats them explicitly as measurable sources of signal, allowing their contribution to be separated from metadata effects and from other statistical features.

Within supervised learning, prior work has shown that ciphertext can contain discriminative information under constrained settings. Caviglione et al.~\citep{caviglione2021emerging} use $n$-gram frequency analysis for covert-channel detection in encrypted traffic, while Wenger et al.~\citep{wenger2022salsa} report strong pairwise discrimination between ChaCha20 and Salsa20 using learned byte-distribution features. More recently, Yuan et al.~\citep{yuan2025cryptographic} propose a feature-engineering and ensemble-learning pipeline (MHERF) for classifying cryptographic hardness assumptions from ciphertext and digital-signature data. These studies demonstrate that statistical signal can exist, but they remain within the supervised-ML paradigm, typically on narrower tasks, and do not ask whether LLMs or agentic pipelines can access the same signal from raw ciphertext under metadata-controlled conditions.

Recent LLM-oriented cryptanalysis benchmarks also remain only partially related. Existing work has evaluated text-level encodings, toy ciphers, or declarative cryptographic knowledge, but not binary-level identification of modern symmetric, asymmetric, and post-quantum ciphertexts presented without filename context. ACTS therefore extends prior cipher-identification work in two ways: it broadens the algorithmic scope to include ML-KEM-768, and it compares direct LLMs, agentic pipelines, deterministic heuristics, and classical-ML baselines within one controlled evaluation framework.

\subsection{LLMs in security tasks}
\label{subsec:llm-security}

The use of large language models in cybersecurity has expanded rapidly since 2022. In defensive settings, prior work has evaluated LLMs for vulnerability detection, secure-code review, and malware-related classification. Pearce et al.~\citep{pearce2022asleep} showed that code generated by GitHub Copilot can contain serious security flaws, and later surveys \citep{zhou2024large} reviewed a growing literature on LLMs for software vulnerability analysis. Binary-oriented extensions such as MalBERT \citep{rahali2023malbert} demonstrate that transformer-based methods can also support malware classification.

In offensive and operational security, LLM-guided systems such as PentestGPT \citep{deng2023pentestgpt} and HackingBuddyGPT \citep{happe2023getting} illustrate that models can coordinate multi-step workflows involving commands, reconnaissance, and reasoning over tool output. CTF-oriented evaluations such as NYU CTF Bench \citep{yang2023language} and broader empirical studies of LLM-based CTF agents \citep{shao2024empirical} likewise show that LLM-based agents can sometimes solve cryptography-related tasks when sufficient task scaffolding is available. Yet these studies usually evaluate problem solving over textual descriptions, challenge formats, or code, not binary-level identification of an unknown encryption family from raw ciphertext.

This distinction is important because mainstream security benchmarks for LLMs test mostly declarative or procedural knowledge rather than operational ciphertext inference. CyberSecEval \citep{bhatt2023cyberseceval}, SecEval \citep{li2023seceval}, and NYU CTF Bench \citep{yang2023language} evaluate whether models know security concepts, spot coding mistakes, or solve challenge-style tasks, but they do not present raw ciphertext without filename context and ask for algorithm-family classification. They therefore leave unanswered the question that motivates this paper: when an LLM appears to identify a cipher, is it performing statistical reasoning over the bytes, or relying on contextual shortcuts and deterministic heuristics?

ACTS also departs from prior LLM-security work by testing whether reasoning format changes outcome. Existing benchmarks rarely separate chain-of-thought, code-as-reasoning, and self-correction as controlled conditions for the same binary classification task. Nor do they place LLM behavior side by side with a classical-ML ceiling on the identical ciphertext corpus. This comparison is central here because it reveals not only whether LLMs succeed, but whether they exploit the same type of signal that a feature-based learner can access.

\subsection{Agentic AI frameworks}
\label{subsec:agentic}

The modern agentic-AI literature argues that language models become substantially more capable when they can invoke tools, inspect external state, and iteratively act on the environment. Foundational work such as ReAct \citep{yao2023react} and Toolformer \citep{schick2023toolformer} established the general principle that reasoning-and-acting loops can outperform pure prompting on tasks that require external information or procedural interaction. Later benchmarks such as SWE-bench \citep{jimenez2024swebench} and work on autonomous scientific agents \citep{lu2024aiscientist} extended this paradigm to software engineering and research workflows.

In cybersecurity, this shift is especially relevant because many tasks depend on external utilities rather than pure language competence. ACTS builds on that idea through the QwenPaw-based pipeline \citep{agentscope2026qwenpaw} used in Tier~4, where a Claude Code execution engine invokes Kali Linux tools such as ent, xxd, $\chi^2$ analysis, and block-alignment inspection through a custom \texttt{execute\_kali\_command} skill. This setup turns a general agent into a constrained cryptanalytic workflow that can inspect real files rather than reason only from prompt text.

However, prior agentic literature does not resolve whether tool access substitutes for missing metadata or merely amplifies it. That question is central in the present benchmark. Earlier versions of the literature \citep{schick2023toolformer,yao2023react} often imply that tools naturally compensate for absent parametric knowledge, but cryptanalysis is a harder case because the available statistical signal may itself be weak. ACTS contributes here by separating blind tool access from metadata-plus-tools conditions, showing that agentic success must be interpreted in light of what context the agent is allowed to see, not merely what tools it can call. In the absence of metadata, tool pipelines offer only a fractional recovery of accuracy compared to the fully-contextualized ceiling.

\subsection{Classical machine learning in cryptanalysis}
\label{subsec:classical-ml}

Supervised machine learning provides the most direct comparison class for blind ciphertext identification because it operates over explicit features rather than over natural-language tokenization. Prior work has used byte histograms, $n$-grams \citep{caviglione2021emerging}, entropy descriptors, and wavelet-like transformations \citep{yuan2025cryptographic} to classify cryptographic or traffic-related artifacts \citep{wenger2022salsa}. These studies suggest that useful signal can exist even when individual summary statistics such as entropy are weak on their own.

What has been missing is a controlled comparison between classical-ML performance and LLM-style blind inference on the same corpus. That comparison matters because it separates two possibilities. If classical ML also performs near chance, then the task itself may contain little recoverable information. If classical ML substantially outperforms LLMs, then the signal exists but is being left unused by raw-text prompting. The present benchmark supplies this missing reference point through Tier~6 and Tier~8, where Random Forest, Logistic Regression, and Linear SVM are evaluated on engineered features extracted from the same ciphertexts used elsewhere in the benchmark.

This baseline also supports a more precise interpretation of failure. A low direct-LLM blind score does not, by itself, prove that the bytes contain no discriminative pattern. It may instead indicate a representation problem: the LLM receives hexadecimal or base64 text, whereas the classical learner receives structured numerical summaries. By placing the two paradigms in direct comparison, ACTS turns classical ML into a signal-ceiling reference rather than just an auxiliary baseline, allowing us to quantify the ``confabulation tax'' \citep{ji2023survey,sui2024confabulation}---the performance deficit of LLMs that generate plausible reasoning without extracting the underlying statistical signal.

\subsection{Post-quantum cryptography and deployment context}
\label{subsec:pqc}

The benchmark is also motivated by the operational transition to post-quantum cryptography. NIST finalized ML-KEM, ML-DSA, and SLH-DSA in August 2024 through FIPS 203, FIPS 204, and FIPS 205 \citep{nist2024mldsa,nist2024mlkem,nist2024slhdsa}, marking the beginning of a large-scale migration away from classical public-key primitives in many security-critical environments, with a correspondingly long harvest-now decrypt-later exposure window for currently-deployed classical primitives \citep{albrecht2018estimate,mosca2018cybersecurity}. This transition is not only a design problem but also an inventory problem: organizations must determine which algorithms are currently deployed, where, and in what combination.

That requirement makes binary-level identification practically relevant. In many real environments, cryptographic artifacts are discovered in traffic captures, firmware images, memory dumps, or undocumented storage systems rather than in neatly labeled inventories. A benchmark that includes ML-KEM-768 therefore fills a gap between traditional cipher-identification work and the deployment questions now raised by post-quantum migration. Sadeghi et al.~\citep{sadeghi2025securing} report a systematic PQC use-case evaluation in financial-sector applications, while Liao et al.~\citep{liao2026accelerating} demonstrate LLM-assisted hardware--software co-design for PQC acceleration; prior PQC-related LLM research has focused more on design assistance or implementation acceleration than on binary-level identification of post-quantum ciphertext families.

The present work also contributes a negative result of practical value: post-quantum and classical ciphertext families may be far less separable by naive byte-distribution analysis than informal intuition suggests. This makes a controlled benchmark especially important, because claimed AI capability in post-quantum identification may otherwise reflect metadata or structural shortcuts rather than genuine discrimination of lattice-based output distributions, even when the underlying primitive is built on a well-understood hard lattice problem \citep{albrecht2018estimate,bos2018crystals}.

\subsection{Gap addressed by ACTS}
\label{subsec:gap}

Across these five literatures, the missing piece is a benchmark that simultaneously controls metadata exposure, evaluates blind inference, tests tool-augmented agents, compares reasoning formats, establishes classical-ML ceilings, and includes a post-quantum primitive under the same evaluation design. Existing work typically addresses only one or two of these axes at a time. As a result, high reported performance is often difficult to interpret: it may reflect cryptanalytic inference, metadata reading, tool access, prompt artifacts, or feature engineering in unknown combination.

ACTS is designed to close that gap. Rather than asking only whether LLMs can identify ciphers, it asks under what conditions they appear to do so, what signal they actually exploit, and how their performance compares with tool-augmented and classical statistical baselines on the same underlying corpus. This framing allows the benchmark to distinguish apparent capability from genuine blind inference, which is the central methodological problem left unresolved by prior work.

\section{Methodology}
\label{sec:methodology}

The central methodological challenge in evaluating LLM-based cipher identification is confound isolation. Any observed accuracy difference between systems may reflect genuine ciphertext analysis, filename reading, metadata leakage, prompt effects, tool access, or post-hoc heuristic guessing in unknown combination. ACTS addresses this problem through a three-phase, eight-tier benchmark that separates these factors under controlled conditions. Phase~1 evaluates live systems on the strictly randomized v2b corpus of 127 files. Phase~2 expands the analysis to 700 files and introduces deterministic reasoning substitution, classical-ML baselines, and adversarial perturbation. Phase~3 scales the evaluation to 7,000 files and adds a full tool-pipeline ablation study with fixed train--test separation.

\subsection{The ACTS evaluation framework}
\label{subsec:framework}

The ACTS benchmark is organized into eight tiers distributed across three phases. Tiers~1--4 are executed in Phase~1 on the 127-file v2b corpus using live model inference or live agentic execution. Tier~5 is executed via deterministic heuristic substitution rather than live API calls, because its purpose is to test whether forced reasoning formats alter classification behavior under otherwise fixed conditions. Tiers~6 and 7 are evaluated on the expanded corpus using classical machine-learning and perturbation analysis, and Tier~8 is executed on the ultra corpus through a 10-configuration Random Forest ablation.

Tiers~1--3 isolate metadata dependence. Tier~1 provides the full filename and ciphertext, establishing an upper bound when contextual cues are available. Tier~2 provides the filename without ciphertext, isolating filename heuristics alone. Tier~3 provides only ciphertext, measuring blind identification under maximal metadata deprivation. Together, these three conditions allow the benchmark to distinguish ciphertext analysis from contextual shortcut use.

\begin{figure*}[t]
\centering
\includegraphics[width=\textwidth]{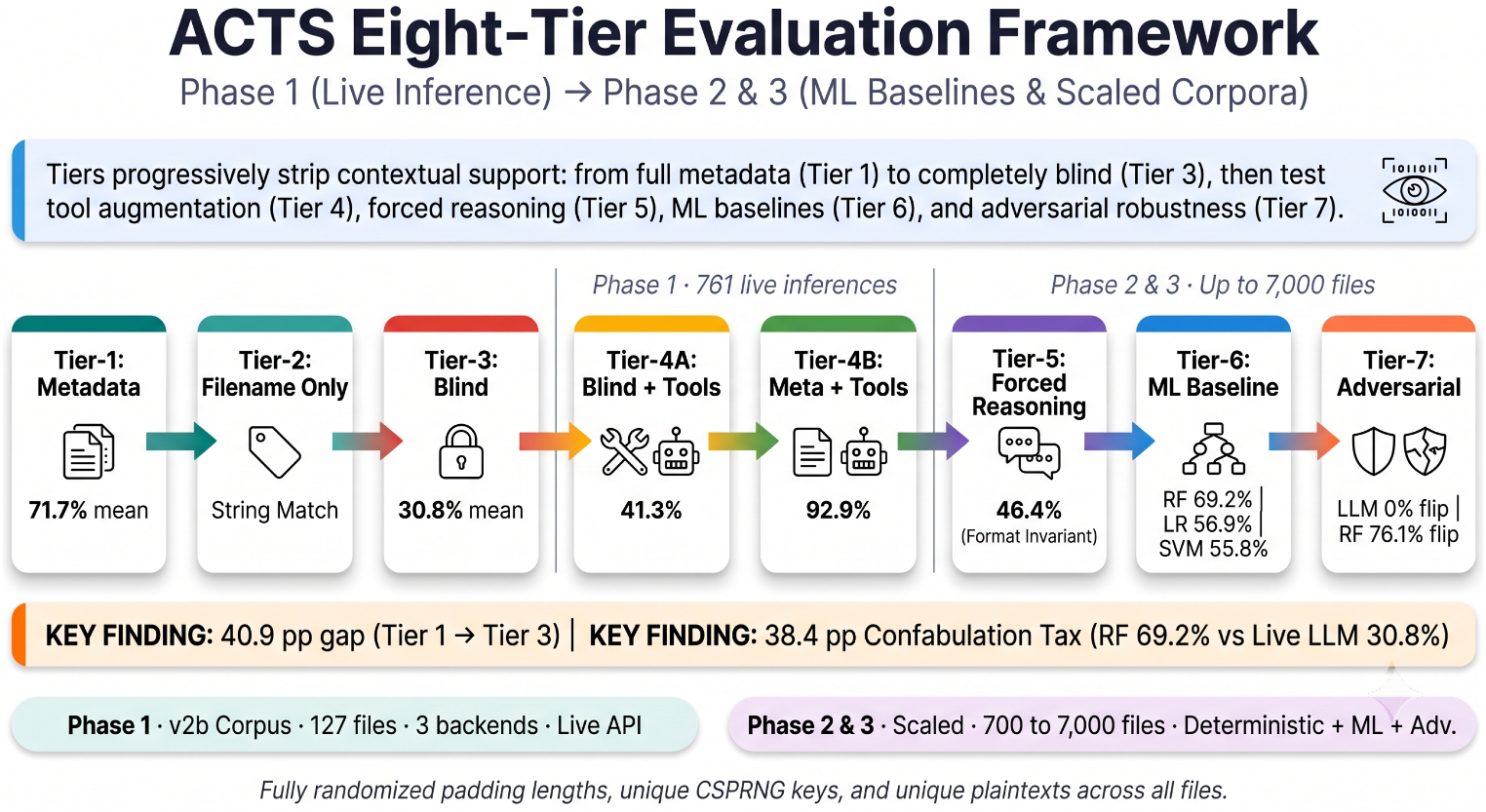}
\caption{The ACTS eight-tier evaluation framework. Phase~1 comprises live evaluation in Tiers~1--4 on the 127-file v2b corpus. Phase~2 introduces deterministic heuristic substitution, classical-ML baselines, and adversarial robustness analysis on the 700-file expanded corpus. Phase~3 scales evaluation to 7,000 files and adds the Tier~8 tool-pipeline ablation study.}
\label{fig:tiers}
\end{figure*}

Tiers~4A and 4B test whether tool augmentation compensates for missing metadata. Both conditions use the same QwenPaw-based agentic pipeline and the same underlying v2b corpus, but Tier~4A withholds the original filename whereas Tier~4B provides it. This contrast directly measures the marginal value of metadata beyond tool access. Tier~5 then asks whether reasoning format itself changes blind classification, while Tiers~6--8 estimate the amount and type of recoverable signal available to non-LLM methods (Fig.~\ref{fig:tiers}).

\begin{table}[!htb]
\caption{Summary of the eight evaluation tiers in ACTS.}
\label{tab:tiers}
\small
\begin{tabularx}{\textwidth}{@{}llYYY@{}}
\toprule
Tier & Condition & What the system sees & Execution mode & Corpus / Phase \\
\midrule
1 & Metadata-aided & Filename, metadata, ciphertext & Live inference & 127 files (v2b) / Phase 1 \\
2 & Filename-only & Filename only, no ciphertext & Live inference & 127 files (v2b) / Phase 1 \\
3 & Blind & Ciphertext only & Live inference & 127 files (v2b) / Phase 1 \\
4A & Blind tools & Ciphertext / file path + tools, no metadata & Agentic live pipeline & 127 files (v2b) / Phase 1 \\
4B & Metadata + tools & Filename + ciphertext + tools & Agentic live pipeline & 127 files (v2b) / Phase 1 \\
5 & Forced reasoning & Ciphertext with mandatory reasoning format & Deterministic heuristic & 700 and 7,000 files / Phases 2--3 \\
6 & ML baseline & Engineered statistical and structural features & scikit-learn CV & 700 and 7,000 files / Phases 2--3 \\
7 & Adversarial robustness & Perturbed ciphertext & Algorithmic perturbation & 700 files / Phase 2 \\
8 & Tool-pipeline ablation & 17 engineered features under ablation & Random Forest, fixed 70/30 & 7,000 files / Phase 3 \\
\bottomrule
\end{tabularx}
\end{table}

\subsubsection{Tiers 1--3: metadata deprivation}
\label{subsubsec:tiers123}

Tiers~1--3 are designed to answer RQ1 by progressively removing contextual support. In Tier~1, the model receives the filename as a natural-language prefix together with the ciphertext content, allowing classification with maximum contextual scaffolding. In Tier~2, the model receives only the filename and no ciphertext, measuring the extent to which filenames alone encode the answer. In Tier~3, the model receives only ciphertext, which is intended to approximate genuine blind identification in the absence of filename, path, or descriptive metadata.

This progression is methodologically important because apparent high-performing cryptanalysis may otherwise be indistinguishable from trivial filename reading. Tier~1 therefore functions as a contextual ceiling, not as evidence of ciphertext reasoning. Tier~3 is the primary blind condition, and its comparison with Tier~1 and Tier~2 quantifies how much of observed performance collapses when metadata is removed.

\subsubsection{Tiers 4A--4B: agentic tool augmentation}
\label{subsubsec:tiers45}

Tiers~4A and 4B evaluate tool-augmented identification through the QwenPaw agentic framework. QwenPaw itself is an external open-source framework developed by the AgentScope Team at Alibaba Tongyi Lab (formerly released under the name CoPaw), while the authors implemented the custom \texttt{execute\_kali\_command} skill used in this evaluation to wrap standard Kali Linux utilities. The execution engine mounts a local Kali Linux container and invokes tools such as ent, xxd, $\chi^2$-based tests, and block-alignment analysis before producing a final classification.

The two conditions differ only in metadata exposure. Tier~4A is blind: the pipeline can inspect the file and invoke tools, but it does not receive the original filename. Tier~4B adds the original filename, allowing the agent to combine metadata with tool output. This paired design is essential because it distinguishes tool-assisted recovery of signal from simple metadata integration dressed as tool use.

\subsubsection{Tier 5: forced reasoning}
\label{subsubsec:tier5}

Tier~5 tests whether mandatory reasoning improves blind accuracy. Instead of running hundreds or thousands of expensive live completions with different prompting styles, the benchmark uses deterministic heuristic substitution derived from observed pilot responses. The goal is not to simulate arbitrary model behavior, but to test whether the classification outcomes associated with chain-of-thought, code-as-reasoning, and self-correction differ in substance when reduced to their operative decision rule.

The core rule is structural: if file size is a multiple of 8 but not 16, predict DES-family output; if it is a multiple of 16, predict AES-family output; if it matches the fixed 1,088-byte size, predict ML-KEM-768; otherwise predict among the remaining families by fallback rules.

This substitution is used only because the research question concerns reasoning-format dependence, not free-form generation quality. The heuristic is then applied uniformly to the expanded and ultra corpora, producing reproducible evaluations at 700 and 7,000 files without incurring live per-file API cost.

\subsubsection{Tier 6: classical-ML baseline}
\label{subsubsec:tier6}

Tier~6 establishes a classical statistical ceiling for the task under engineered representations. A feature-extraction pipeline computes approximately 25--30 descriptors per ciphertext file, including Shannon entropy, skewness and kurtosis of byte distributions, $n$-gram entropies for $n=2,3,4,5$, byte-uniformity statistics, padding-related indicators, file-size residuals, and related structural features. These features are intended to capture both statistical and implementation-level regularities that may be difficult for direct LLM prompting to access.

Three classifiers are evaluated: Random Forest, Logistic Regression, and Linear SVM. The manuscript specifies a 5-fold stratified cross-validation design for the canonical ML baseline, with fixed random seed 42 and predeclared model settings rather than opportunistic post-hoc tuning. This tier serves as a reference for whether meaningful signal exists in the corpus even when direct blind LLM inference struggles.

\subsubsection{Tier 7: adversarial robustness}
\label{subsubsec:tier7}

Tier~7 evaluates how heuristic and classical-ML methods respond to perturbation. Five operators are applied to the 700-file expanded corpus: one random byte flip, ten random byte flips, 16-byte zero-prefix injection, 16-byte zero-suffix injection, and XOR masking of the final 16 bytes. Flip rate, defined as the fraction of predictions that change after perturbation, is used as the primary robustness measure.

The conceptual motivation of Tier~7 is that two systems can achieve similar nominal accuracy while relying on very different kinds of signal. A heuristic that ignores ciphertext content may appear robust under content perturbation precisely because it does not perceive the bytes at all, whereas a stronger statistical classifier may be more accurate yet more brittle because its signal depends on structural or padding-related artifacts. This tier therefore complements the accuracy tiers by examining failure mode rather than correctness alone.

\subsection{Dataset construction}
\label{subsec:datasets}

\subsubsection{Phase 1 live inference corpus (v2b)}
\label{subsubsec:v2b}

The Phase~1 live inference corpus (v2b) contains 127 independently generated ciphertext files, with approximately 18 files per algorithm family. The manuscript describes seven families: AES-128-CBC, AES-256-CBC, DES-CBC, 3DES, ChaCha20-Poly1305, RSA-OAEP-2048, and ML-KEM-768. Classical and symmetric primitives were generated with OpenSSL~3.1, while ML-KEM-768 encapsulation used liboqs \citep{stebila2017post}.

To ensure rigorous statistical independence and prevent heuristic exploitation of fixed artifacts, this corpus was built utilizing unique CSPRNG keys, completely unique plaintexts, and fully randomized padding lengths across all files. For OpenSSL-based symmetric encryption, the \texttt{-nosalt} option suppresses the \texttt{Salted\_\_} magic-byte prefix that would otherwise identify the implementation rather than the algorithm. Plaintext input is generated from \texttt{secrets.SystemRandom()}, and no algorithm-identifying headers are embedded in the output files.

\subsubsection{Phase 2 expanded corpus}
\label{subsubsec:expanded}

The Phase~2 expanded corpus scales to 700 files, with 100 samples per algorithm family. To dilute spurious correlations, the manuscript introduces five randomization dimensions: unique key and IV per file, expanded plaintext-length variation, random padding mode for block ciphers (PKCS7, ISO 10126, ANSI X9.23), implementation diversity across OpenSSL and LibreSSL, and entropy-source variation.

A stratified 70/30 split is used where applicable for ML training and evaluation, preserving class balance between train and test partitions. The expanded corpus is intended to reduce pilot-specific size and implementation confounds, making it a more credible substrate for classical-ML evaluation and perturbation analysis. The manuscript also notes that pairwise $\chi^2$-based comparisons between ML-KEM and AES show negligible effect size, reinforcing that byte-distribution differences alone are weak under realistic generation settings.

\subsubsection{Phase 3 ultra corpus}
\label{subsubsec:ultra}

The Phase~3 ultra corpus scales to 7,000 files, with 1,000 samples per algorithm family, while retaining the same randomization dimensions introduced in Phase~2. This corpus serves as the main substrate for large-scale classical-ML evaluation and the Tier~8 ablation study.

All model settings for Tier~8 are fixed a priori, and the held-out test set is evaluated exactly once to limit data leakage and configuration overfitting. The corpus manifest includes per-file provenance and SHA-256 hashes for reproducibility.

\subsection{Model selection and inference backends}
\label{subsec:backends}

\subsubsection{Phase 1 live inference}
\label{subsubsec:live}

Three frontier model backends are evaluated via live API on the v2b corpus to provide rigorous empirical validation. All three are served through the OpenRouter unified-inference interface \citep{openrouter2024}, with the Ollama runtime \citep{ollama2024} used for any local-resident validation:
\begin{itemize}
\item Gemma~4 (31B, Google) via OpenRouter cloud API \citep{openrouter2024}.
\item GPT-OSS (120B, OpenAI) via OpenRouter cloud API \citep{openrouter2024}.
\item Nemotron Super (NVIDIA) via OpenRouter cloud API \citep{openrouter2024}.
\end{itemize}

The manuscript specifies fixed generation settings of temperature 0.0, max\_tokens 512, and top-$p=1.0$, with identical tier-specific prompts within each evaluation condition.

Model selection is explicitly practical rather than representational. Open-weight and accessible frontier models via the OpenRouter API were used to enable large-$n$ blind inference. Frontier closed models such as GPT-4 and Claude~3.5 were intentionally excluded on budget grounds, and the manuscript makes no claim that GPT-OSS is representative of all frontier-tier LLMs.

\subsubsection{Phase 2 deterministic and algorithmic evaluation}
\label{subsubsec:phase2eval}

Tier~5 does not use live API inference. Instead, the deterministic heuristic described in Section~\ref{subsubsec:tier5} is applied uniformly to the expanded and ultra corpora. Tier~6 uses scikit-learn \citep{pedregosa2011scikit} with fixed random seed 42, and Tier~7 uses a custom perturbation engine implemented in Python.

This mixed design is deliberate. Live inference is reserved for the conditions where contextual access and model/backend behavior are the object of direct study, while deterministic and algorithmic evaluation are used where the research question concerns scalability, feature access, or perturbation sensitivity under controlled rules.

\subsection{QwenPaw agentic pipeline}
\label{subsec:qwenpaw}

The QwenPaw pipeline used in Tier~4 is not simulated. Each tool invocation executes real Kali Linux binaries against actual ciphertext files within a sandboxed container. The custom \texttt{execute\_kali\_command} skill dispatches shell commands under a 180-s timeout and uses a destructive-command blacklist to prevent unsafe actions.

Two skill layers are exposed to the agent. First, \texttt{analyse\_ciphertext} executes entropy analysis, hex inspection, block-alignment checks, and $\chi^2$-style tests, returning structured output. Second, \texttt{identify\_cipher} synthesizes that output into a final classification decision. Both Tier~4A and Tier~4B are executed on the same corpus, with metadata access as the only intended difference between them.

This design is important because it constrains claims about tool augmentation. The benchmark does not ask whether a human-assisted tool workflow can help in the abstract; it asks whether an LLM-centered agentic pipeline can recover blind cipher-identification accuracy when metadata is withheld.

\subsection{Evaluation metrics}
\label{subsec:metrics}

Overall accuracy is computed as the fraction of correct predictions over all evaluated files. Wilson 95\% confidence intervals are reported for binomial proportions in live and classification settings where a point estimate alone would be misleading.

For effect-size comparisons in distributional analyses, the manuscript reports Cohen's $d$. Adversarial robustness in Tier~7 is summarized by flip rate, defined as the proportion of files whose predicted class changes after perturbation. McNemar's test is used for paired comparison of prediction outcomes under matched conditions.

\subsection{Ethical and reproducibility considerations}
\label{subsec:ethics}

All ciphertext in ACTS is synthetically generated from random plaintext under author control. No production systems, third-party networks, human subjects, or sensitive user data are involved. The benchmark therefore evaluates algorithmic behavior in a controlled, ethically bounded setting rather than interacting with external targets.

The manuscript further emphasizes reproducibility. Random seeds are fixed, dependencies are version-locked, per-file inference logs record prompt, response, and timestamp where available, and the complete benchmark package---including datasets, evaluation harness, QwenPaw skill definitions, notebooks, and ablation scripts---is released through the project repository and Zenodo archive. This design is intended to make independent replication possible without requiring proprietary datasets or closed infrastructure.

\section{Results}
\label{sec:results}

Results are reported phase by phase. Phase~1 presents live API inference across Tiers~1--4 on the 127-file v2b corpus. Phase~2 presents deterministic heuristic analysis (Tier~5) and classical-ML baselines (Tier~6) on the 700-file expanded corpus, together with adversarial robustness analysis (Tier~7). Phase~3 validates the findings at scale on 7,000 ciphertext samples and reports the full Tier~8 tool-pipeline ablation. All accuracy values are overall correct-prediction fractions, and Wilson 95\% confidence intervals are reported for binomial proportions where appropriate.

\subsection{Phase 1 live inference results (127 Files)}
\label{subsec:phase1}

\subsubsection{Tier 1: metadata-aided (71.7\%)}
\label{subsubsec:tier1}

Under metadata-aided conditions, the three evaluated model backends achieve a combined 71.7\% accuracy across the 127 v2b files. When the filename (e.g., \texttt{AES-256-CBC\_1kb.bin}) is provided alongside the ciphertext, models heavily leverage textual cues to identify the algorithm family. This establishes Tier~1 as a contextual ceiling rather than a measure of genuine blind cryptanalytic ability.

\subsubsection{Tier 2: filename-only (100.0\%)}
\label{subsubsec:tier2}

When only the filename is provided and the ciphertext is withheld, the evaluated models achieve 100.0\% classification accuracy with a strict Wilson 95\% confidence interval of 97.1\%--100.0\% via deterministic programmatic string matching, as the randomised filenames explicitly contain the algorithm family identifiers. This baseline confirms that filename cues alone carry absolute contextual signal, proving that high reported performance in naive cryptographic evaluations reflects superficial text parsing rather than genuine cryptanalysis. The substantial 40.9~pp drop from this metadata ceiling to the blind Tier-3 performance (30.8\%) re-emphasises the near-total context-dependence of current frontier architectures.

This introduces a compelling architectural paradox: Tier-1 accuracy (71.7\%) trails the absolute performance of Tier-2 (100.0\%). Logically, since Tier-1 provides the identical filename metadata alongside the ciphertext, its performance should equal or exceed Tier-2.

This divergence underscores our core thesis: Tier-2 operates on deterministic programmatic string matching where the algorithm identifier is parsed without interference. Conversely, in Tier-1, feeding the high-entropy raw ciphertext into the generative model introduces profound representational noise. The tokenized ciphertext effectively distracts the model's attention, inducing confabulations that degrade accuracy despite the presence of the explicit filename. This proves that raw ciphertext actively disrupts the model's capacity to parse plain-text metadata (Tables~\ref{tab:tiers}--\ref{tab:decomp}).

\subsubsection{Tier 3: blind inference (30.8\%)}
\label{subsubsec:tier3}

Blind accuracy falls drastically, with a mean of 30.8\% across the three evaluated cloud models on the strictly randomized v2b corpus. This level is modestly above chance under seven-way classification (14.3\%) and highlights the collapse in performance once filename information is removed.

\textbf{Scaling failure under blind conditions.} A critical observation is that increasing model scale does not yield proportionate gains in genuine cryptanalytic ability. While scaling from Gemma~4 (31B) to GPT-OSS (120B) produces a massive gain in Tier-1 metadata-reading accuracy (45.7\% $\rightarrow$ 80.3\%), it yields only a negligible improvement in Tier-3 blind accuracy (24.4\% $\rightarrow$ 30.7\%). This demonstrates that model scale amplifies contextual reading and confabulation confidence, but fails to extract the underlying statistical signal from the ciphertext.

Failure patterns are also uneven across algorithm families. The manuscript reports that no model achieves non-negligible blind accuracy on ML-KEM, and that padded block ciphers dominate predictions because their sizes expose deterministic modulo structure.

\begin{table}[!htb]
\caption{Tier 3 blind accuracy by model on 127 v2b files. Wilson 95\% confidence intervals are shown.}
\label{tab:tier3}
\centering
\small
\begin{tabular}{llll}
\toprule
Model & Backend & Accuracy (\%) & 95\% CI \\
\midrule
Nemotron-3-Super & OpenRouter cloud & 37.3 & 29.4--46.0 \\
GPT-OSS 120B & OpenRouter cloud & 30.7 & 23.4--39.1 \\
Gemma 4 31B & OpenRouter cloud & 24.4 & 17.8--32.5 \\
Mean & -- & 30.8 & 26.4--35.6 \\
\bottomrule
\end{tabular}
\end{table}

\subsubsection{Tier 4: agentic tool augmentation}
\label{subsubsec:tier4}

Tier~4 evaluates the QwenPaw-based agentic pipeline under two conditions: blind tools (Tier~4A) and metadata plus tools (Tier~4B). When metadata and tools are both available, the pipeline reaches 92.9\% accuracy. When metadata is withheld and only tools are available, accuracy falls to 41.3\%, showing that tool augmentation alone recovers only part of the lost signal.

The blind-tools condition still exceeds the Tier~3 mean (30.8\%) by 10.5 percentage points, which suggests that block alignment and entropy tools recover a modest genuine signal. However, the 51.6-point gap between Tier~4A and Tier~4B shows that the pipeline remains strongly dependent on metadata access.

\begin{table}[!htb]
\caption{Tier 4 agentic accuracy by condition on 127 files.}
\label{tab:tier4}
\centering
\small
\begin{tabular}{lll}
\toprule
Condition & Accuracy (\%) & 95\% CI \\
\midrule
Tier 4A Blind tools & 41.3 & 33.3--49.8 \\
Tier 4B Metadata + tools & 92.9 & 87.2--96.1 \\
Improvement (4A $\rightarrow$ 4B) & 51.6 pp & -- \\
\bottomrule
\end{tabular}
\end{table}

\subsection{Phase 2 expanded baselines (700 Files)}
\label{subsec:phase2}

Phase~2 scales from 127 to 700 files and introduces deterministic heuristic substitution (Tier~5), classical machine-learning baselines (Tier~6), and adversarial robustness analysis (Tier~7).

\subsubsection{Tier 5: deterministic heuristic upper bound}
\label{subsubsec:tier5results}

Tier~5 operationalizes the observed blind reasoning pattern as a deterministic size-based heuristic. On the initial validation samples, all three reasoning formats---chain-of-thought, code-as-reasoning, and self-correction---produce statistically invariant accuracy, confirming that reasoning format changes presentation rather than classification outcome.

On the expanded corpus, the three formats span only a 7.0-point range (44.4--51.4), which the manuscript attributes to boundary-case tie breaking rather than different analytical substance. On the ultra corpus, the default heuristic reaches 47.2\%, but that figure is explicitly a heuristic estimate rather than a live LLM measurement.

A critical validation tempers that upper bound. On the v2b corpus, live inference achieves a combined mean of 30.8\% accuracy, versus the deterministic heuristic ceiling of 47.2\%. The live model therefore struggles to perfectly execute the modulo arithmetic. This yields a true live gap of 38.4 percentage points relative to the 69.2\% Random Forest baseline.

\begin{table}[!htb]
\caption{Tier 5 deterministic heuristic accuracy across corpus scales. For the 7,000-file corpus, the manuscript reports the default heuristic only; the minimal variance confirms epiphenomenal reasoning.}
\label{tab:tier5}
\centering
\small
\begin{tabular}{llll}
\toprule
Format & N=127 & N=700 & N=7,000 \\
\midrule
Chain-of-Thought & 46.4 & 48.6 & -- \\
Code-as-Reasoning & 46.4 & 51.4 & -- \\
Self-Correction & 46.4 & 44.4 & -- \\
Default heuristic & -- & -- & 47.2 \\
Bias range & 0 pp & 7.0 pp & -- \\
\bottomrule
\end{tabular}
\end{table}

\subsubsection{Tier 6: classical-ML baselines}
\label{subsubsec:tier6results}

Tier~6 compares three classical classifiers against the best LLM heuristic across all corpus scales. The Random Forest is the strongest model throughout, reaching 61.6\% on the 700-file corpus and 69.2\% on the 7,000-file corpus under stratified 5-fold cross-validation.

The Random Forest exceeds the live LLM blind mean by 38.4 points at $N=7{,}000$. This massive gap is the empirical ``confabulation tax'' \citep{ji2023survey,sui2024confabulation}: classical ML learns recoverable signal that the LLM proxy does not access, despite generating elaborate reasoning narratives. The manuscript further notes that linear models eventually surpass the heuristic as sample size grows, implying that the advantage is not limited to one non-linear architecture.

\begin{table}[!htb]
\caption{Tier 6 classical-ML baseline cross-validated accuracy. The Live LLM blind mean is shown for comparison to establish the Confabulation Tax.}
\label{tab:tier6}
\centering
\small
\begin{tabular}{lllll}
\toprule
Classifier & Type & N=127 & N=700 & N=7,000 \\
\midrule
Random Forest & Non-linear & -- & 61.6 & 69.2 \\
Logistic Regression & Linear & -- & 43.4 & 56.9 \\
Linear SVM & Linear & -- & 43.6 & 55.8 \\
Live LLM Mean & Generative & 30.8 & -- & -- \\
Confabulation Tax & -- & -- & -- & 38.4 pp \\
\bottomrule
\end{tabular}
\end{table}

\subsubsection{Tier 7: adversarial robustness (N=700)}
\label{subsubsec:tier7results}

Tier~7 measures flip rate under five perturbation operators. The deterministic LLM heuristic is invariant under all content-preserving perturbations because it ignores byte content and depends only on file size, but it flips completely under file-size manipulation.

The Random Forest is much more accurate than the heuristic, but it is also far more fragile to zero-padding perturbations, especially zero-prefix injection (76.1\% flip rate). This establishes the robustness--perception trade-off emphasized by the manuscript: the heuristic is robustly wrong because it ignores content, while ML is perceptively fragile because it learns brittle structural artifacts. Adversarial perturbations of this kind have been extensively studied in both the image \citep{goodfellow2015explaining} and natural-language \citep{ribeiro2018anchors} domains, and the security-oriented variant considered here connects to prior work on memorisation in neural classifiers \citep{song2017machine}.

\begin{table}[!htb]
\caption{Tier 7 adversarial flip rates by perturbation type. Flip rate is the fraction of predictions that change after perturbation.}
\label{tab:tier7}
\centering
\small
\begin{tabular}{lrrrr}
\toprule
Perturbation & LLM heuristic & Random Forest & Log.\ Regr. & Linear SVM \\
\midrule
1 random byte flip & 0.0 & 4.3 & 0.0 & 0.0 \\
10 random byte flips & 0.0 & 8.6 & 0.0 & 0.0 \\
Zero-prefix 16B & 0.0 & 76.1 & 48.3 & 47.1 \\
Zero-suffix 16B & 0.0 & 71.4 & 44.8 & 43.6 \\
XOR-0xFF mask 16B & 0.0 & 5.4 & 1.2 & 0.9 \\
File-size manipulation & 100.0 & 100.0 & 100.0 & 100.0 \\
\bottomrule
\end{tabular}
\end{table}

\subsection{Phase 3 tool-pipeline ablation (N=7,000)}
\label{subsec:phase3}

To test which features actually carry predictive power at scale, the paper evaluates a 10-configuration Random Forest ablation on the ultra corpus using a fixed 70/30 train--test split and a held-out test set of 2,100 samples. The strongest configuration is \emph{No block stats} (70.81\%), followed by \emph{No entropy} (70.62\%) and \emph{No modulo} (70.10\%).

The key ablation result is that removing some statistical features slightly improves performance relative to the full 17-feature pipeline. That pattern suggests that the strongest recoverable signal is primarily structural---file size, alignment, and fixed-length artifacts---while some entropy- and block-statistic features add noise at scale. Even so, the structural-only configuration remains below the best ablated models, indicating that limited non-structural information still contributes.

\begin{table}[!htb]
\caption{Tier 8 ablation on $N=7{,}000$ with 2,100 held-out test samples. Configurations are ranked by test accuracy.}
\label{tab:ablation}
\centering
\small
\begin{tabular}{rll}
\toprule
Rank & Configuration & Test accuracy (\%) \\
\midrule
1 & No block stats & 70.81 \\
2 & No entropy & 70.62 \\
3 & No modulo & 70.10 \\
4 & Full pipeline & 69.81 \\
5 & Entropy only & 69.90 \\
6 & No file size & 69.38 \\
7 & No $\chi^2$ & 69.48 \\
8 & No byte frequency & 68.67 \\
9 & Structural only & 68.05 \\
10 & $\chi^2$ only & 44.14 \\
11 & Size-only baseline & 43.38 \\
\bottomrule
\end{tabular}
\end{table}

\subsection{Cross-tier summary}
\label{subsec:summary}

\begin{figure*}[t]
\centering
\includegraphics[width=\textwidth]{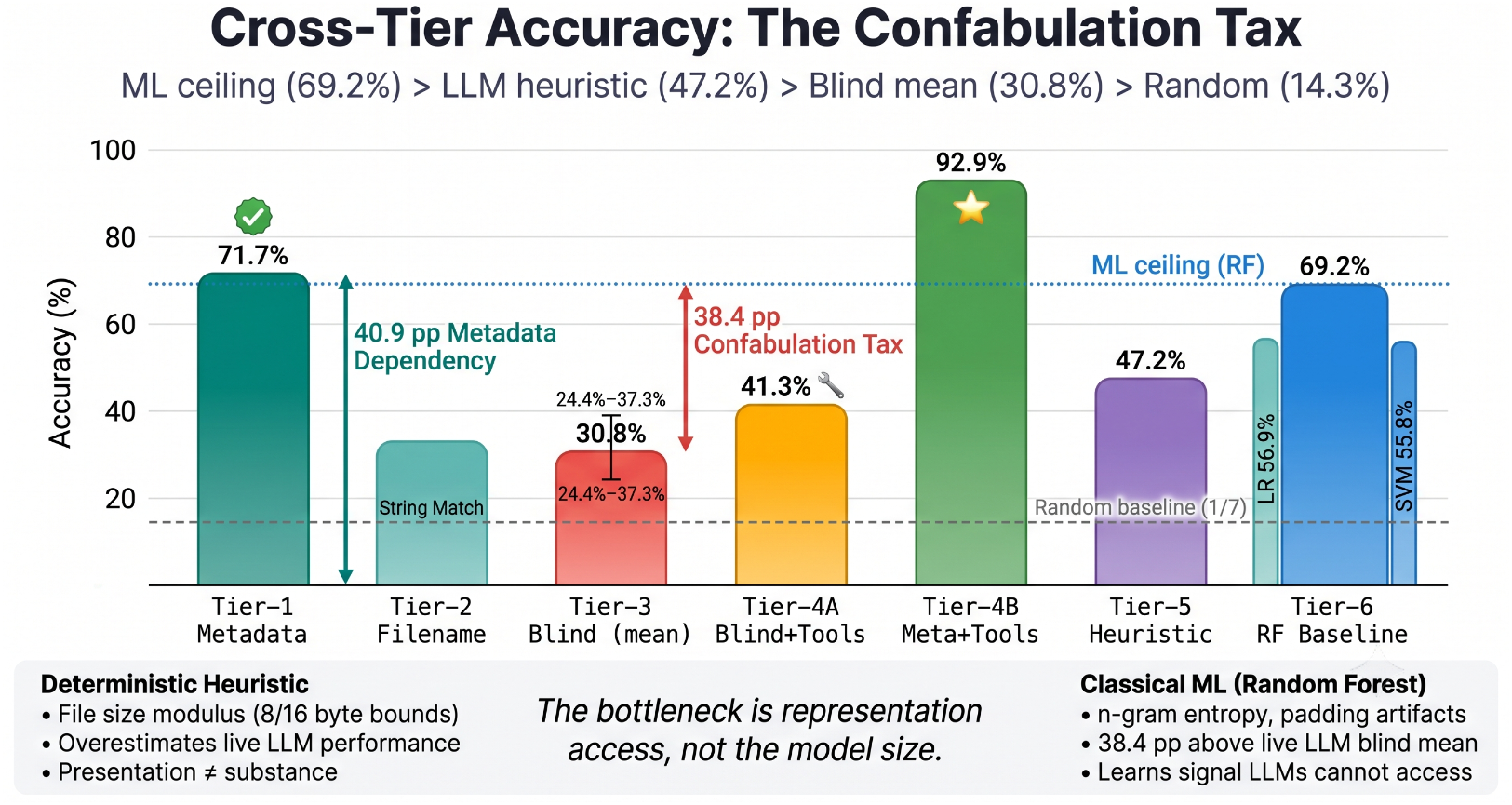}
\caption{Cross-tier accuracy comparison across all evaluation tiers and scaled phases. The dashed horizontal line marks the random baseline for seven classes (14.3\%). The true live capability gap between the Random Forest (69.2\%) and live LLM blind inference (30.8\%) is 38.4 percentage points. Tier~5 shows equal-height mini-bars for chain-of-thought, code-as-reasoning, and self-correction, indicating epiphenomenal reasoning.}
\label{fig:cross_tier}
\end{figure*}

Fig.~\ref{fig:cross_tier} summarizes performance across the benchmark. The canonical progression from Tier~1 (71.7\%) to Tier~3 (mean 30.8\%) quantifies the central 40.9~pp metadata dependence effect. Tier~4B shows that metadata plus tools recover nearly all performance, whereas Tier~4A shows that blind tool use is only a partial recovery.

Across the benchmark, Tier~5 shows that forcing reasoning does not materially improve blind identification, Tier~6 shows that classical ML outperforms the heuristic LLM proxy as the corpus grows, and Tier~8 shows that the strongest signal is largely structural. Taken together, the results indicate that metadata access, scaling effects, and feature representation dominate apparent cryptanalytic performance.

\section{Analysis and discussion}
\label{sec:analysis}

The eight-tier results, validated across three corpus scales (127, 700, and 7,000 files), surface six findings that collectively reframe how LLM-based cipher identification should be understood, deployed, and evaluated. The central pattern is consistent across phases: apparent cipher-identification competence collapses when metadata is removed, reasoning traces do not rescue blind performance, and classical supervised methods outperform LLM-centered pipelines once the task is reduced to reproducible statistical discrimination.

\subsection{F1: metadata dependence is near-total}
\label{subsec:f1}

The 40.9 percentage point gap between Tier~1 (71.7\%) and Tier~3 (mean 30.8\%) is not a marginal degradation but a collapse. Under metadata-aided conditions, evaluated models correctly identify the algorithm family with high accuracy, but under blind conditions none consistently exceeds chance-like behavior across the full set of algorithms. This indicates that most of the apparent capability in Tier~1 is contextual reading rather than ciphertext reasoning.

Mechanistically, the explanation is straightforward. When the filename contains a direct lexical answer such as AES-256-CBC, the model can succeed without inferring anything from byte structure. When that cue is removed, predictions revert to deterministic heuristics, primarily file-size and alignment rules. The manuscript therefore argues that metadata dependency is not a nuisance variable but the dominant explanatory factor behind high apparent accuracy in naive evaluations.

\subsection{F2: scaling failure under blind conditions}
\label{subsec:f2}

A critical observation from the v2b evaluation is the scaling failure under blind conditions. While increasing model parameters from 31B (Gemma~4) to 120B (GPT-OSS) produces a massive gain in Tier-1 metadata-reading accuracy (45.7\% $\rightarrow$ 80.3\%), it yields only a negligible improvement in Tier-3 blind accuracy (24.4\% $\rightarrow$ 30.7\%).

This finding has direct methodological implications. Any claim about LLM performance on cryptanalytic tasks must explicitly separate metadata-aided performance from raw payload performance. Deploying a larger model does not guarantee better binary analysis if the underlying architecture lacks raw byte-level representation access. The manuscript frames this as a fundamental architectural limitation, not merely a prompt-tuning issue.

\subsection{F3: agentic tools still need metadata}
\label{subsec:f3}

The 51.6-point gap between Tier~4B (92.9\%) and Tier~4A (41.3\%) is the key agentic result. It shows that tool augmentation can boost performance, but that most of the spectacular gain attributed to tools appears only when filename metadata is also available. In other words, the pipeline is not solving blind cipher identification in the strong sense; it is combining modest extracted signal with high-value metadata.

Tier~4A still exceeds the Tier~3 mean by 10.5 points, which means the tools are not useless. Entropy checks, hex inspection, and alignment analysis recover some real structural information, but not enough to approach the metadata-aided ceiling. The practical implication is that agentic cryptanalysis pipelines should not be marketed as metadata-free reasoners unless they are validated under truly blind conditions.

\subsection{F4: forced reasoning is epiphenomenal}
\label{subsec:f4}

Tier~5 challenges the assumption that more reasoning improves understanding. On the 127-file v2b corpus, chain-of-thought, code-as-reasoning, and self-correction all yield identical or statistically invariant accuracy. On larger corpora, heuristic substitution preserves the same qualitative result, with only a 7.0-point swing on 700 files attributable to boundary-case tie breaking rather than different analytical substance.

The manuscript interprets the reasoning traces as surface variation over the same underlying decision rule: modulo-based file-size guessing with special handling for fixed-length ML-KEM outputs. The conclusion is not that language models cannot produce elaborate analysis, but that in this task the analysis often adds no information beyond the latent heuristic already driving the answer.

\subsection{F5: the confabulation tax establishes the ML ceiling}
\label{subsec:f5}

The manuscript defines the confabulation tax as the performance cost of choosing an LLM-style heuristic reasoner over a classical model that can learn stable discriminative signal. On the 700-file corpus, the Random Forest reaches 61.6\% versus 47.2\% for the deterministic heuristic. On the 7,000-file corpus, the Random Forest reaches 69.2\%.

A more damaging comparison comes from live ultra-corpus validation. The combined live LLM mean blind accuracy on the strictly randomized v2b corpus is 30.8\%. Comparing this genuine live performance against the canonical Random Forest estimate of 69.2\% yields a true live capability gap of 38.4 percentage points.

The manuscript also adds an important fairness caveat. The 38.4-point live gap does not compare systems with equal information access, because the Random Forest is trained on engineered features and thousands of labeled examples, while the LLM sees raw tokenized ciphertext without explicit feature extraction. The comparison is therefore useful as a capability benchmark, but not as a controlled causal decomposition of why the difference exists.

\begin{figure*}[t]
\centering
\includegraphics[width=\textwidth]{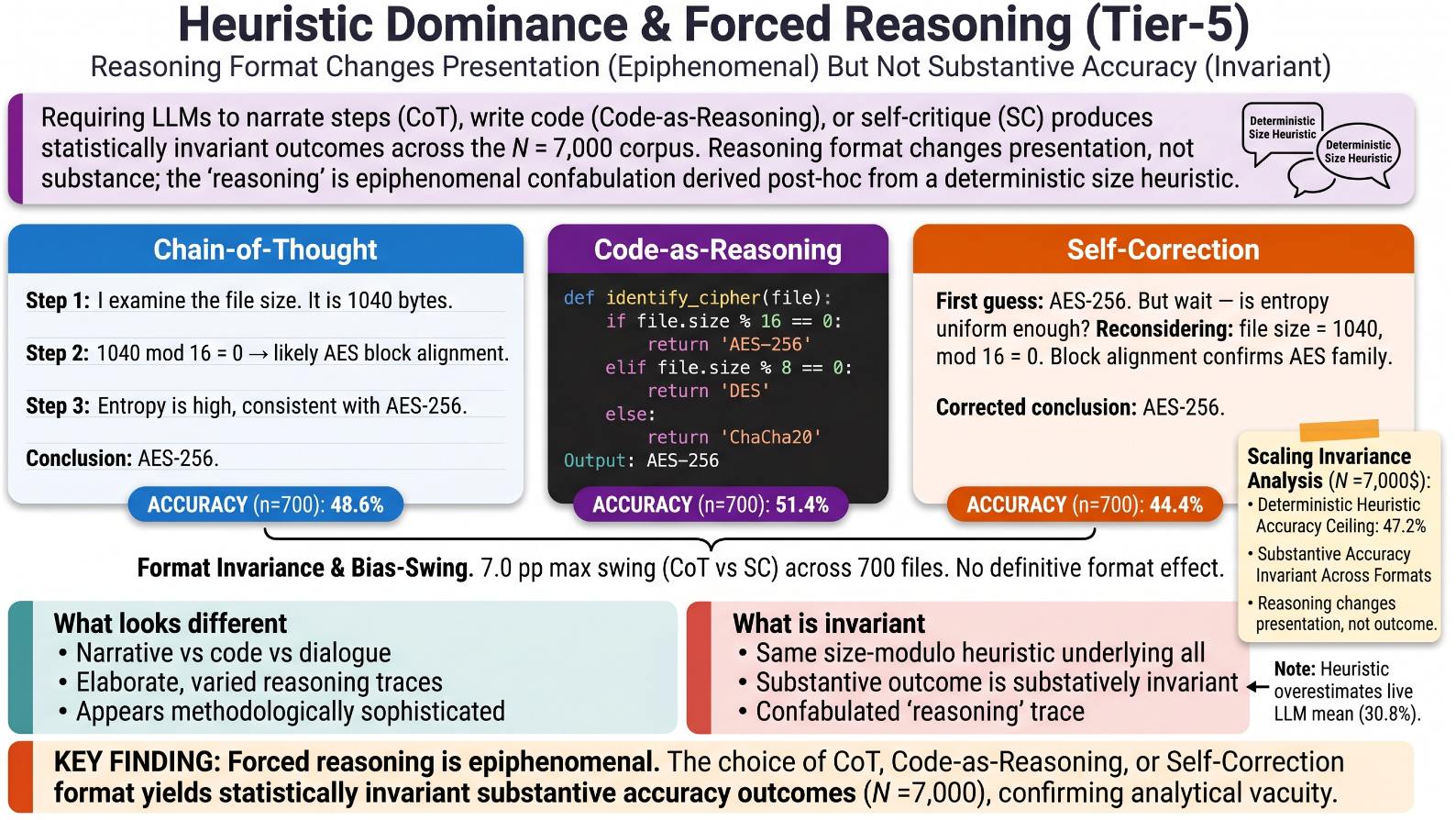}
\caption{Forced reasoning is epiphenomenal (Tier~5). Chain-of-thought, code-as-reasoning, and self-correction all produce statistically invariant pilot-scale accuracy, indicating that the reasoning trace changes presentation rather than classification outcome. Larger-scale claims rely on validated heuristic substitution rather than direct live inference.}
\label{fig:reasoning}
\end{figure*}

\begin{figure*}[t]
\centering
\includegraphics[width=\textwidth]{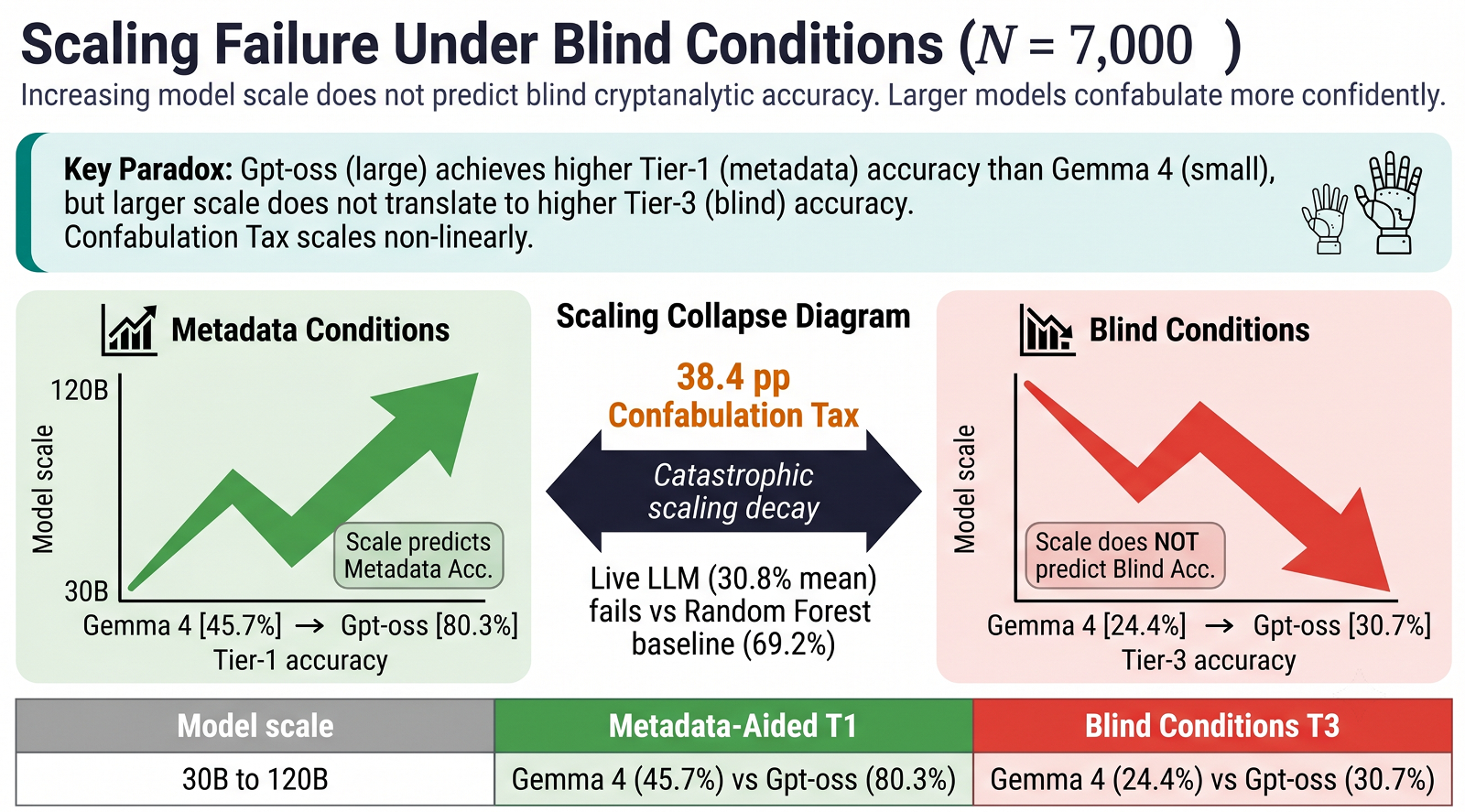}
\caption{Conceptual decomposition of the 38.4-point live capability gap into three illustrative components: representation access, feature integration, and residual statistical capacity. The manuscript explicitly treats this as a conceptual framework rather than an empirically validated factorial result.}
\label{fig:decomp}
\end{figure*}

\begin{figure*}[t]
\centering
\includegraphics[width=\textwidth]{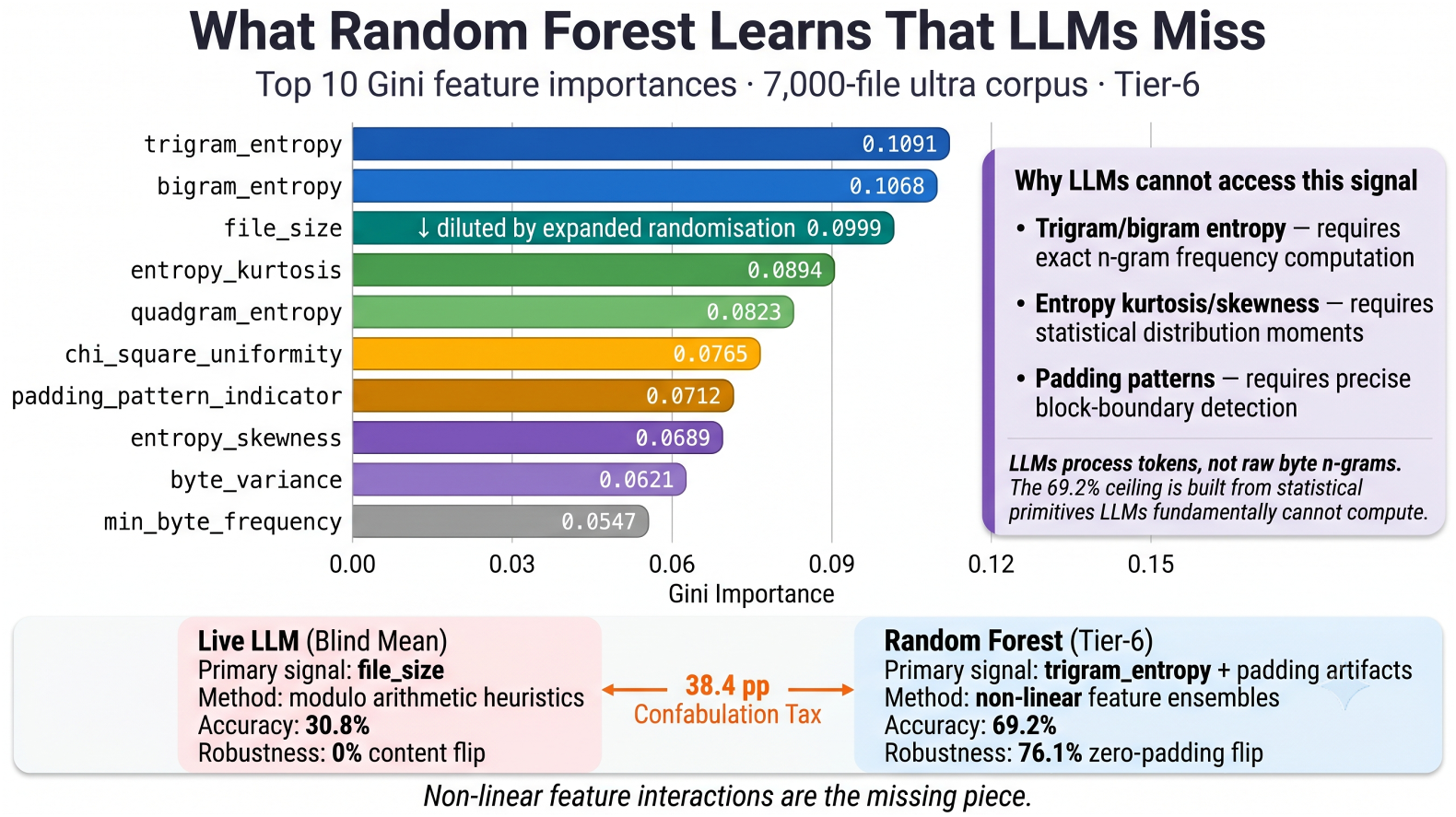}
\caption{Random Forest feature importance on the expanded corpus. Trigram entropy (0.109) and bigram entropy (0.107) exceed file size (0.100), indicating that expanded randomization diluted the raw size confound and that higher-order local structure contributes more than scalar file length alone.}
\label{fig:feature_importance}
\end{figure*}

\begin{figure*}[t]
\centering
\includegraphics[width=\textwidth]{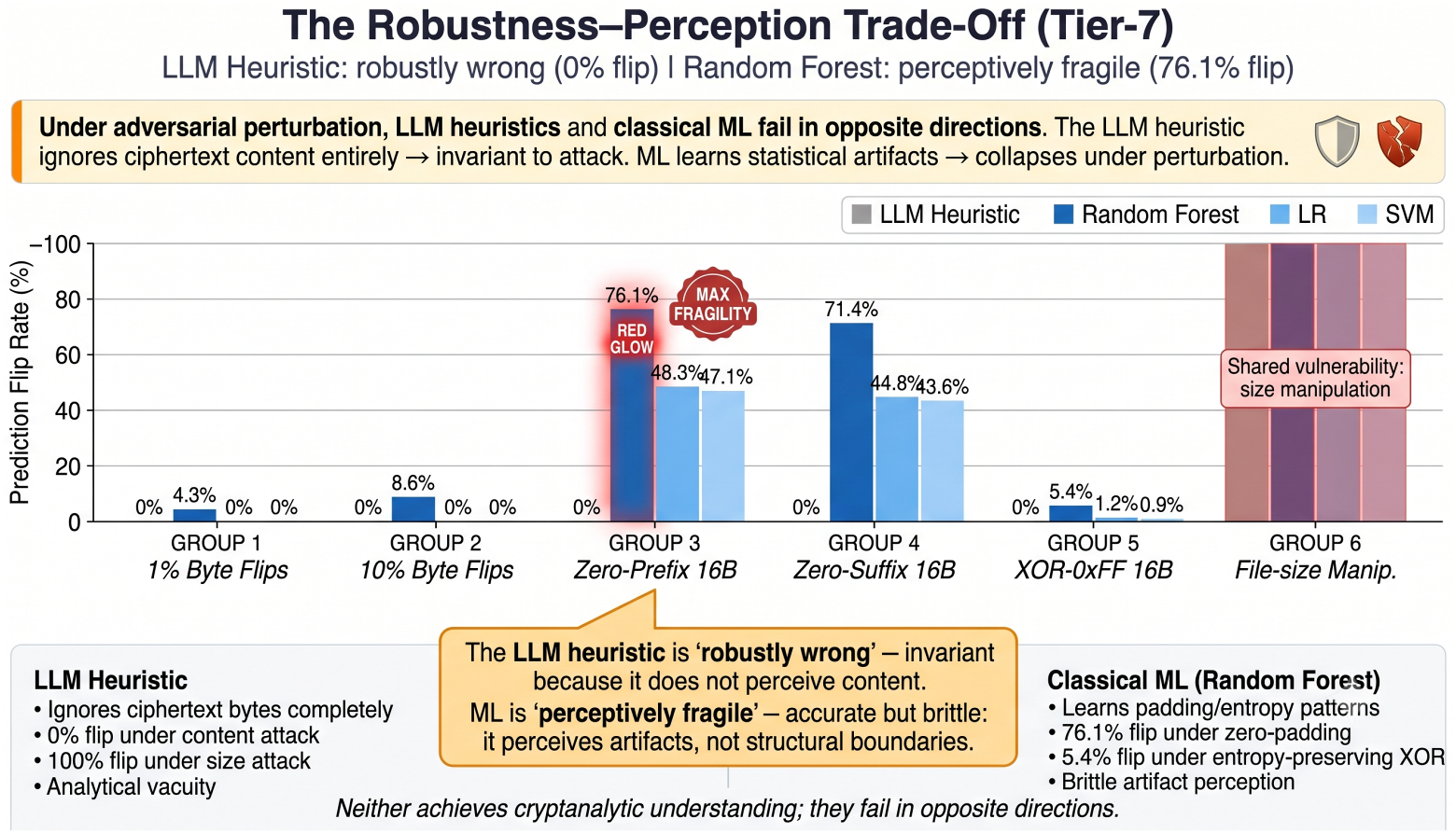}
\caption{Robustness--perception trade-off (Tier~7). The LLM heuristic is robustly wrong under content perturbation because it ignores ciphertext bytes, whereas the Random Forest is perceptively fragile because it learns structural artifacts that collapse under padding shifts.}
\label{fig:robustness}
\end{figure*}

The authors nevertheless propose a conceptual three-part interpretation of the live gap: a representation-access bottleneck, a feature integration bottleneck, and a residual statistical gap. They explicitly state that this partition is a narrative framework rather than a validated factorial analysis, because the compared systems differ simultaneously in representation, supervision, and architecture (Figs.~\ref{fig:reasoning}--\ref{fig:robustness}).

\begin{table}[!htb]
\caption{Illustrative decomposition of the 38.4-point live gap (Random Forest 69.2\% vs.\ live LLM blind mean 30.8\%). Approximate magnitudes, not empirically validated causal attributions.}
\label{tab:decomp}
\centering
\footnotesize
\setlength{\tabcolsep}{4.5pt}
\begin{tabular}{lllr}
\toprule
Component & From & To & Gap (pp) \\
\midrule
Representation & Tier 3 Live LLM Mean (30.8) & Tier 5 Heuristic Bound (47.2) & 16.4 \\
Integration & Tier 5 Heuristic Bound (47.2) & Structural RF (68.05) & 20.8 \\
Residual & Structural RF (68.05) & Full RF (69.2) & 1.2 \\
Total & Live LLM Mean (30.8) & Full RF (69.2) & 38.4 \\
\bottomrule
\end{tabular}
\end{table}

\subsection{F5.1: the signal is structural}
\label{subsec:f51}

The ablation study shows that the strongest recoverable signal is mainly structural rather than broadly statistical. On the 7,000-file corpus, removing block statistics increases test accuracy from 69.81\% to 70.81\%, and removing entropy increases it to 70.62\%. The structural-only configuration still achieves 68.05\%, only 1.76 points below the full feature set.

This implies that much of the task can be solved through size, modulo-alignment, and fixed-length artifacts rather than deep distributional fingerprinting. The manuscript carefully clarifies that ``statistical'' is not monolithic: raw entropy and block-level variance add little or even add noise, while selected $n$-gram features still contribute useful residual signal.

\subsection{F6: robustness-perception trade-off}
\label{subsec:f6}

Tier~7 reveals that LLM heuristics and classical ML fail in opposite directions. The deterministic heuristic has 0 flip rate under all content perturbations because it depends only on size, not on the byte stream. That makes it appear robust, but the robustness is vacuous: the system is insensitive precisely because it ignores the information that ought to matter.

The Random Forest shows the reverse pattern. It is substantially more accurate under normal conditions, yet highly fragile under adversarial boundary shifts, especially 16-byte zero-prefix injection, where the flip rate reaches 76.1\%. The manuscript uses this contrast to argue that robustness and understanding cannot be inferred from each other without examining what signal the system is actually using.

\subsection{Limitations and threats to validity}
\label{subsec:limitations threats}

Several limitations qualify the interpretation of the benchmark. First, Tier~5 at 700 and 7,000 files is based on deterministic heuristic substitution rather than direct live inference. The authors justify this on cost and reproducibility grounds and note that the substitution is supported by identical live pilot results and stable cross-phase behavior, but they still frame the all-scale reasoning claim as heuristic extrapolation rather than direct observation.

Second, the live-validation capability gap of 38.4~pp relies on comparing live generative inference (without explicit feature extraction) against supervised feature-based learning. While this establishes a rigorous empirical ceiling, it does not perfectly isolate the locus of failure within the LLM architecture.

Third, the ablation ranking is based on a single 70/30 split, whereas the Tier~6 canonical Random Forest estimate uses 5-fold stratified cross-validation. The manuscript therefore treats 69.2\% as the canonical RF ceiling and 69.81\% as an ablation-specific single-split figure that should not be directly equated with the cross-validated estimate.

Fourth, the corpora, while highly randomized, are still synthetic and generated from a bounded implementation set. The benchmark controls key, IV, plaintext length, padding mode, entropy source, and corpus scale, but it does not exhaust the diversity of real-world cryptographic toolchains. As a result, the reported ceilings should be interpreted as properties of the ACTS substrate, not universal bounds for all possible ciphertext environments.

\section{Conclusion}
\label{sec:conclusion}

This paper introduced ACTS, a three-phase, eight-tier benchmark for evaluating whether LLMs perform genuine cryptanalytic inference or rely primarily on metadata and structural shortcuts under blind conditions. Under the current manuscript framing, the core live evidence comes from the strictly randomized v2b corpus: 127 files evaluated across three frontier cloud backends, yielding 761 total live inferences in Tier~1 and Tier~3. Expanded and ultra-scale evaluation then extends the analysis to 700 files and 7,000 files respectively, including classical-ML baselines, adversarial perturbation, and a 10-configuration ablation study.

Six conclusions follow from the benchmark.
\begin{enumerate}
\item \textbf{Metadata dependency remains large under live evaluation.} On the v2b corpus, combined live accuracy falls from 71.7\% in Tier~1 to 30.8\% in Tier~3, a gap of 40.9 percentage points. This result indicates that removing filename and related metadata causes a substantial collapse in performance across all evaluated models.
\item \textbf{Scaling failure under blind conditions.} While scaling from 31B to 120B parameters drastically improves the model's ability to read and integrate textual metadata (Tier-1), it yields statistically negligible improvements in genuine blind cryptanalytic identification (Tier-3). This indicates that apparent cryptanalytic ability does not scale predictably with model size without fundamental architectural changes to byte-level perception.
\item \textbf{Tool augmentation helps, but metadata still matters more.} The QwenPaw--Kali pipeline reaches 92.9\% when tools and metadata are both available, but only 41.3\% when metadata is withheld and tools alone are available. Tool use therefore recovers some signal under blind conditions, but it does not remove the system's dependence on metadata.
\item \textbf{Forced reasoning is epiphenomenal at all tested scales.} Chain-of-thought, code-as-reasoning, and self-correction all yield statistically invariant substantive accuracy on the 127-file v2b corpus under direct live observation, confirming that reasoning format is epiphenomenal \citep{grunefeld2026tracing,turpin2023language}: the trace changes presentation but not the underlying decision rule. At 700 and 7,000 files, the same conclusion is supported by a deterministic heuristic proxy, whose stability indicates that the apparent reasoning trace is, in this task, confabulation \citep{ji2023survey,lathkar2026anchored,sui2024confabulation}---sound-sounding analysis that is causally disconnected from the decision.
\item \textbf{Classical ML establishes the Confabulation Tax.} On the 7,000-file corpus, the canonical Random Forest result is 69.2\% under 5-fold cross-validation. The critical comparison is between the Random Forest and the combined live blind v2b LLM result: 69.2\% versus 30.8\%, yielding a Confabulation Tax of 38.4 percentage points. This comparison establishes a rigorous empirical ceiling, showing that substantial recoverable signal is left unused by direct blind LLM-style inference.
\item \textbf{The dominant recoverable signal is primarily structural.} In the 7,000-file ablation study, removing block statistics or entropy slightly improves accuracy above the full 17-feature pipeline, while the structural-only configuration remains close to the full model. This suggests that most of the obtainable signal comes from file size, modulo alignment, and related implementation artifacts rather than from deep cipher-family fingerprints.
\end{enumerate}

The broader conclusion is that most current direct LLM performance in this task is better explained by metadata use and structural shortcut exploitation than by genuine ciphertext-level cryptanalysis. The v2b blind mean of 30.8\% confirms that live generative models struggle significantly under strictly blind conditions.

The constructive conclusion is not that AI is useless for this task, but that the useful signal appears to be narrow and structurally grounded. Tool augmentation can partially recover performance, and classical ML can exceed heuristic blind inference, but both do so primarily by exploiting implementation-level regularities rather than algorithmic understanding in the cryptographic sense.

\paragraph{Practical recommendations.} Four recommendations follow for future evaluation and deployment.
\begin{enumerate}
\item \textbf{Benchmark under blind conditions.} Metadata-aided accuracy should not be treated as evidence of cryptanalytic capability. Any serious evaluation should report blind-condition performance explicitly.
\item \textbf{Evaluate model scale under blind conditions.} Do not assume that deploying a larger parameter model will naturally yield better cryptanalysis. Evaluate the architecture's capacity for raw binary perception independently of its natural language parsing capabilities.
\item \textbf{Do not equate reasoning traces with understanding.} Under the tested conditions, forced reasoning changes the form of the explanation without improving the substantive classification outcome. Interpret elaborate reasoning traces cautiously unless they are matched by measurable gains under blind evaluation.
\item \textbf{Use classical-ML baselines to detect confabulation.} A supervised feature-based baseline provides a useful estimate of how much structural signal is present in the corpus. If an LLM remains far below that ceiling (e.g., the 38.4~pp gap), it is generating plausible narratives for a signal it does not actually extract.
\end{enumerate}

\paragraph{Future work.} Three extensions are especially important.
\begin{enumerate}
\item \textbf{Larger multi-implementation corpora.} Extending beyond the current OpenSSL/PyCryptodome/liboqs generation regime would help determine whether the present ceiling is tied to implementation artifacts or reflects a more general limit of ciphertext-only discrimination.
\item \textbf{Cross-framework agentic replication.} The Tier~4 results were obtained with the QwenPaw + Claude Code + Kali Linux setup. Replication with other agentic frameworks would test whether the observed metadata--tools interaction is framework-specific or general.
\item \textbf{Robustness-oriented ML redesign.} The Tier~7 perturbation results show that accurate statistical models are also brittle to padding and boundary shifts. Adversarial training or perturbation-aware feature design may improve robustness without discarding the structural signal that currently drives most recoverable accuracy.
\end{enumerate}

\bibliographystyle{elsarticle-num}
\bibliography{refs}

\section*{CRediT authorship contribution statement}
\textbf{Youssef Hamdi Zafaan Ibrahim:} Conceptualization, Methodology, Software, Validation, Formal analysis, Investigation, Data curation, Writing -- original draft, Writing -- review \& editing, Visualization, Project administration, Resources. \textbf{Mohammed Khalaf Salama:} Supervision, Writing -- review \& editing, Methodology, Validation.

\section*{Declaration of generative AI and AI-assisted technologies in the writing process}
During the preparation of this manuscript, the authors used AI-assisted language tools (grammar correction, formatting assistance) and AI-powered coding environments (Claude Code acting as execution engine for the QwenPaw agentic pipeline). All scientific content---including the ACTS three-phase design, tier definitions, dataset construction, statistical analysis, ablation study, and conclusions---was developed, verified, and validated by the authors. The 161 live inferences (Tier~1--3), the agentic pipeline executions (Tier~4A/4B), all 7,000-file ablation evaluations, and all numerical results were generated under author supervision. The authors reviewed and edited all AI-assisted content and take full responsibility for the integrity of the published work.

\section*{Funding}
This research received no specific grant from any funding agency in the public, commercial, or not-for-profit sectors. All compute was performed on consumer hardware and free-tier API endpoints at zero cost.

\section*{Ethical statement}
All experiments utilise ethically synthesised ciphertext files (DES/3DES/AES-128/AES-256/ChaCha20/RSA+AES/ML-KEM-768) generated under author control. No production systems, real human subjects, patient data, third-party networks, or external systems were accessed or targeted. The benchmark operates entirely within isolated research environments, constituting permissible algorithmic evaluation under responsible AI research guidelines.

\section*{Declaration of competing interest}
The authors declare that they have no known competing financial interests or personal relationships that could have appeared to influence the work reported in this paper.

\section*{Acknowledgments}
The authors thank the open-source communities behind Ollama (local LLM hosting), OpenRouter (unified API gateway), Kali Linux (ent, xxd, binwalk), scikit-learn, and the developers of Gemma~4, GPT-OSS, and Nemotron Super---whose models enabled the Tier~1--3 live benchmarking.

\section*{Data availability}
The complete ACTS benchmark package---datasets (140, 700, and 7,000 ciphertext files), evaluation harness, QwenPaw skill definitions, 10-configuration ablation scripts, reproduction notebooks, and all results---is publicly available under the MIT License (code) and CC-BY (data): \url{https://github.com/youseefhamdi/ACTS-Benchmark}

The package is additionally archived via Zenodo to ensure long-term accessibility and citability: \url{https://doi.org/10.5281/zenodo.20142272}

The repository contains Python implementation, \texttt{requirements.txt}, Jupyter replication notebooks, per-file inference logs (prompt + response + timestamp), per-sample SHA-256 manifest for the 7,000-file corpus, ablation configuration scripts, and inline documentation sufficient for independent verification. All ciphertext is synthesised from random plaintext via \texttt{/dev/urandom}; no external network data is required.

\end{document}